\documentclass[10pt,twocolumn]{paper}
\usepackage{epsfig, graphicx}
\usepackage[english]{babel}
\usepackage[margin=1.5cm]{geometry}

\title{\center \rm \bf Modelless X-Ray Scattering Study of a Silica Hydrosol Surface}

\author{\small \rm Aleksey M. Tikhonov$^a$\/\thanks{tikhonov@kapitza.ras.ru}, Viktor E. Asadchikov$^b$,
Yurii O. Volkov$^b$, Boris S. Roshchin$^b$, Veijo Honkim\"aki$^c$, and Maria V. Blanco$^c$\\
\small $^a$Kapitza Institute for Physical Problems, Russian Academy of Sciences, Moscow, 119334 Russia\\
\small $^b$Shubnikov Institute of Crystallography, Federal Research Center Crystallography and Photonics,\\
\small Russian Academy of Sciences, Moscow, 119333 Russia\\
\small $^c$European Synchrotron Radiation Facility, 71 Avenue des Martyrs, 38000 Grenoble, France}

\begin{document}
\maketitle

\abstract{ \it The structure of the adsorbed layer of alkali ions on the surface of colloidal silica solutions with a particle size of 27\,nm has been studied by reflectometry and diffuse scattering of synchrotron radiation with a photon energy of about 71\,keV. Electron density profiles in the direction perpendicular to the surface have been reconstructed from experimental data and spectra of the correlation function of heights in the surface plane have been obtained. The revealed deviation of the integral and frequency characteristics of the roughness spectra of the silica sol surface from predictions of the capillary-wave theory is of a fundamental character. This deviation is due to the contribution from roughnesses with low spatial frequencies $\nu < 10^{-4}$\,nm$^{–1}$ and to the interference of diffuse scattering from different layer interfaces of the surface structure.}

\vspace{0.25in}

The surface of a colloidal solution of SiO$_2$ nanoparticles stabilized by alkali metal hydroxide is
strongly polarized because of the difference between the potentials of "electric image" forces for univalent metal cations and particles with a large negative charge (macroions) [1]. It was reported in [2] that heavy Cs$^+$ ions are selectively collected at the hydrosol–air interface, removing light Na$^+$ ions. The dependence of the electrostatic free energy of a charged sphere at the water–air interface between two dielectric media on its radius $r$ obtained in [3,4] makes it possible to explain the substitution of Cs$^+$ cations with a larger radius $r\approx 1.8$\,\AA{} for Na$^+$ ions with a smaller radius $r\approx 1.2$\,\AA{} [5].

In this work, we report new experimental data on diffuse scattering under the conditions of total external
reflection and synchrotron radiation reflectometry with a photon energy of about 71\,keV for the surfaces
of both NaOH-stabilized and CsOH-enriched silica sols. In contrast to the previous studies, the data are
analyzed within a self-consistent approach that makes it possible to reconstruct both the electron density
profiles perpendicular to the hydrosol surface and spectra of the height-height correlation function in the surface plane without any prior information on the surface structure [6,7].

Figure 1 shows a qualitative four-layer model of the structure of the hydrosol–air interface proposed in [2,8]. This model is based on the structural parameters extracted from synchrotron radiation X-ray reflectometry data with a photon energy of about 15\,keV. This model involves a standard model approach (see, e.g., [9-11]) with plotting electron density profiles based on the error function used in the capillary-wave theory developed by Buff, Lovett, and Stillinger [12]. The total width of the surface transition layer is about the Debye screening length in the solution ($> 200$\,{\AA} at pH < 10). The first layer ($\sim 8$\,{\AA} thick) is a layer of "suspended" metal ions with an estimated surface density of $\sim 4\cdot10^{18}$\,m$^{-2}$. The second layer ($\sim 13$\,{\AA} thick) is a layer of the spatial charge of hydrated Na$^+$ ions with the estimated surface density of sodium ions of $\sim 8\cdot 10^{18}$\,m$^{-2}$, which is independent of the presence of heavy ions in the bulk of hydrosol. Depleted layer 3 with a low density of the electrolyte ($\sim 10$\,nm thick) separates the first two layers from negatively charged particles in the fourth layer. Finally, the thickness of layer 4 coincides with the diameter of colloidal nanoparticles in the solution and their surface density is much higher than their density in the bulk of the solution. This model is in agreement with small-angle grazing scattering and diffraction data [1,13].

\begin{figure}
\epsfig{file=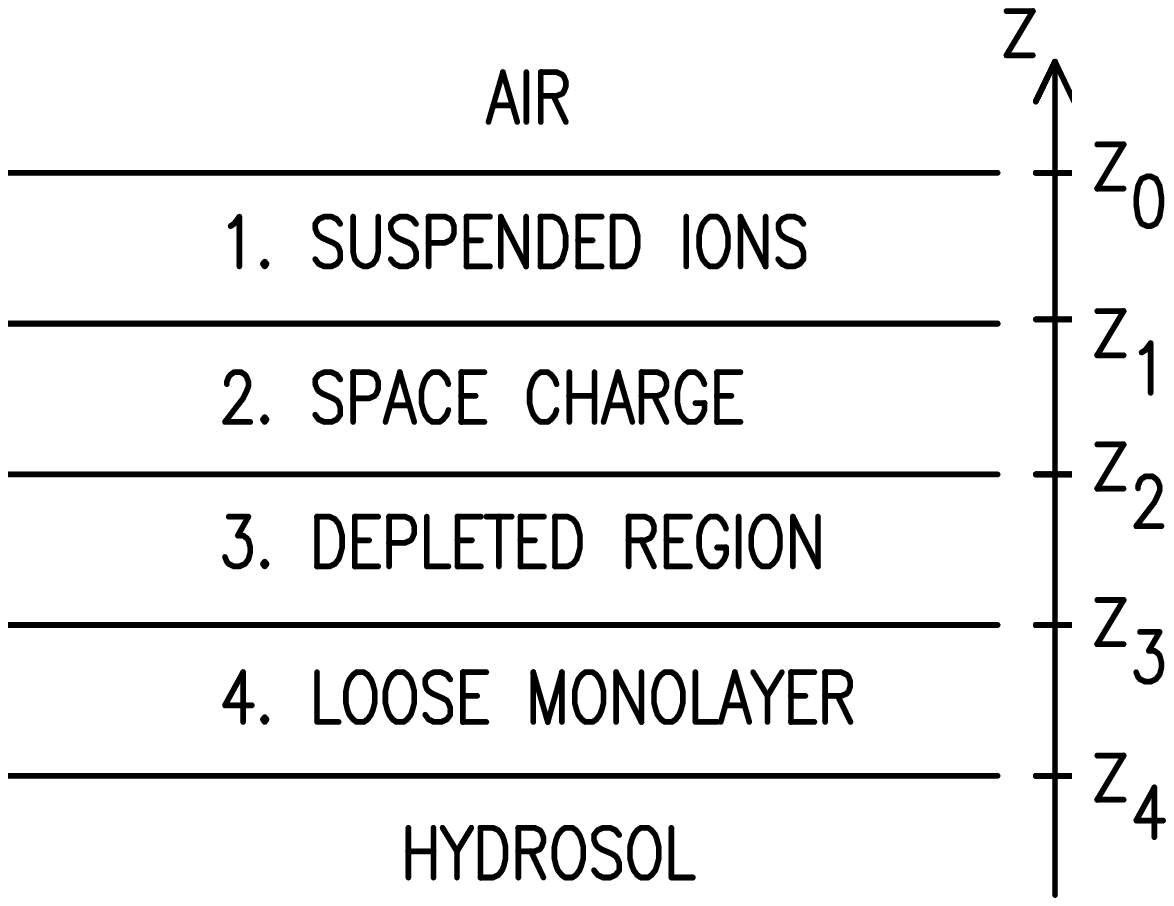, width=0.45\textwidth}

\small {\bf Figure 1.}  {\it Four-layer model of the transverse section of the transition layer at the (silica hydrosol-air) interface [2, 8]. Alkali metal ions with a surface density of $\sim 10^{19}$\,m$^{–2}$ are located in two layers: layer 1 of suspended ions with a low density and a thickness of $\sim 8$\,{\AA}  and layer 2 of hydrated ions with a thickness of $\sim 13$\,{\AA}. Layer 3 with a low density of the electrolyte has a thickness of $\sim 10$\,nm and the thickness of monolayer of nanoparticles (layer 4) is determined by their diameter.}
\end{figure}

\begin{figure}
\epsfig{file=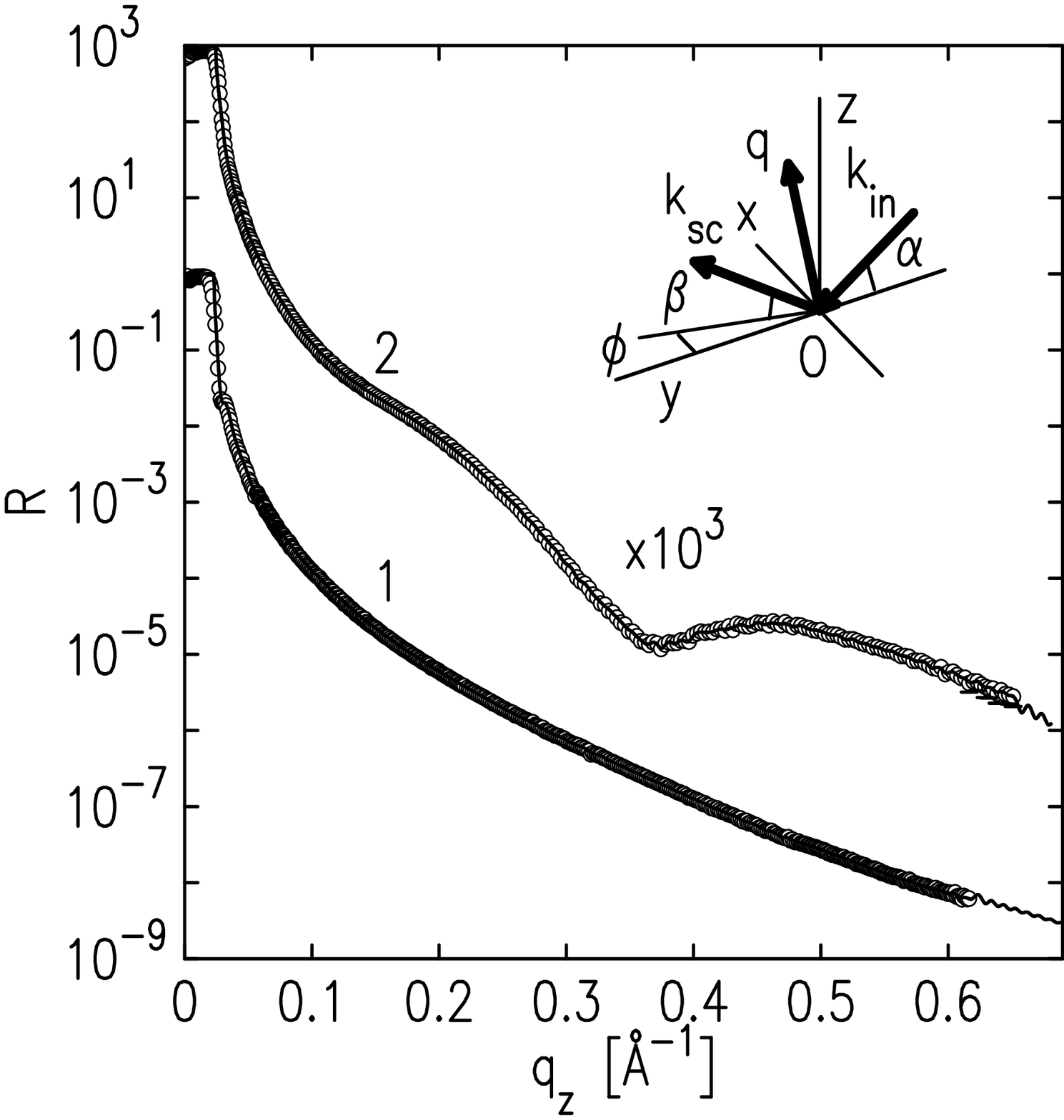, width=0.4\textwidth}

\small {\bf Figure 2.} {\it Reflectivity curves $R(q_z)$ obtained under normal conditions for the surface of (1) NaOH-stabilized and (2) Cs-enriched silica sols. The inset shows the kinematics of
scattering on the liquid surface.}
\end{figure}

The silica sol samples were prepared and studied in an airtight cell with X-ray-transparent windows
according to the method described in [8]. The initial concentrated monodisperse Ludox TM-50 sol stabilized
by sodium hydroxide was purchased from the Grace Davison Co. (pH = 9, 50 wt \% SiO$_2$ and 0.2 wt \%
Na). Further, the sol was either diluted by deionized water (ELGA, PURELAB Option-Q) or enriched by
mixing in a vessel (shaken and then placed in a Bandelin ultrasonic bath) with a solution of cesium
hydroxide in deionized water to a solution with $\approx 30$ wt \% of SiO$_2$. The concentration of Na$^+$ ions in the solutions was $\sim 0.06$\,mol/L (pH = 9) and the concentration of Cs$^+$ in the enriched solution was $\sim 0.6$\,mol/L (pH = 12). Solid hydrate CsOH\,$\cdot$\,x(H$_2$O) (99.9 wt \% metal and 15–20 wt \% H2O) was purchased from the Alfa Aesar Co.

According to the small-angle scattering data, the initial Ludox TM-50 suspension contains homogeneous
amorphous silica particles with a characteristic diameter of $\sim 27$\,nm [14]. Adding CsOH to the initial
hydrosol stabilized by NaOH (pH\,$ \sim 9$), one can obtain a solution with a high volume concentration of Cs$^+$ (pH < 12), which remains liquid in the airtight cell at room temperature at least for a month [15, 16]. At a very high concentration of cesium hydroxide (pH > 12.5), Ludox sols usually become opaque and are
solidified into a gel in about a week. In this process, the size distribution of SiO$_2$ particles hardly changes [17].

The reflectivity $R$ and intensity of surface diffuse Xray scattering $I_d$ at the hydrosol–air interface were measured under normal conditions on the ID31 beamline of the ESRF synchrotron [18]. In the experiments, the intensity of the focused monochromatic photon beam with a wavelength of $\lambda=0.1747 \pm 0.0003$\,\AA{} (photon energy of $\approx 71$\,keV) was $\sim 10^{10}$\,photon/s at transverse dimensions of $\sim 10$\,$\mu$m in height and $\sim 250$\,$\mu$m in the horizontal plane. We already used reflectometry and diffuse scattering data jointly to determine the structures of the liquid–air and liquid–
liquid interfaces but within the model approach [19-21].

It is convenient to describe the kinematics of grazing scattering at the macroscopically flat interface oriented by the gravitational force in the coordinate system with the origin $O$ at the center of the illuminated region. The $xy$ plane coincides with the interface between the monolayer and water, the axis $Ox$ is perpendicular to the beam direction, and the axis $Oz$ is perpendicular to the surface and is opposite to the gravitational force (see the inset of Fig. 2). Let {\bf k}$_{\rm in}$ and {\bf k}$_{\rm sc}$ be the wave vectors with the amplitude $k_0= 2\pi/\lambda$ of the incident and scattered beams in the direction to the observation point, respectively; $\alpha$ be the glancing angle in the $yz$ plane; $\beta$ be the scattering angle such that $\alpha, \beta << 1$; and $\phi$ be the angle between the incident beam and scattering direction in the $xy$ plane. Thus, the components of the scattering vector {\bf q = k$_{\rm in}$ {\rm -} k$_{\rm sc}$} in the interface plane are $q_x=k_0\cos\beta\sin\phi$ and $q_y=k_0(\cos\beta\cos\phi-\cos\alpha)$ and the projection on the axis $Oz$ is $q_z=k_0(\sin\alpha+\sin\beta)$.

The surface samples were prepared and studied at $T=298$\,K in a Teflon dish with a diameter of 100\,mm
placed in an airtight single-stage thermostat. The total external reflection angle $\alpha_c=\lambda\sqrt{r_e\rho_b/\pi}$ $\approx 3\cdot10^{-4}$\,rad (where $r_e =2.814\cdot10^{-5}$\,\AA  is the classical radius of the electron) for the sol–air interfaces is determined by the bulk electron density $\rho_b\approx 1.2 \rho_w$ in solutions, where $\rho_w=0.333$\,e$^{-}$/\AA$^{3}$ is the electron density in water under normal conditions.

At specular reflection ($\alpha = \beta$, $\phi=0$), the scattering vector {\bf q }  is directed along the $Oz$ axis, i.e., $q=q_z\approx 2k_0\alpha$. Figure 2 shows curves $R(q_z)$ for the surface of (1) NaOH-stabilized and (2) Cs-enriched sols. At $q_z < q_c=(4\pi/\lambda)\alpha_c$$\approx 0.025$\,\AA$^{-1}$, the incident beam undergoes total external reflection; i.e., $R\approx 1$.

Figure 3 shows the two-dimensional map of the surface scattering intensity for hydrosol stabilized by
NaOH as a function of the angles $\beta$ and $\phi$ at the glancing angle $\alpha \approx 2.1 \cdot 10^{-4}$\,rad ($\approx 0.012^\circ$). Figure 4 shows
the data for the surface diffuse (nonspecular) scattering intensity $I_d (\beta)$ for the surface of (1) NaOH-stabilized and (2) CsOH-enriched sols at the same $\alpha$  value and $\phi = 0$ (see the inset of Fig. 4). The most intense peak on these curves corresponds to the specular reflection at $\beta \approx 0.7 \alpha_c$.

\begin{figure}
\epsfig{file=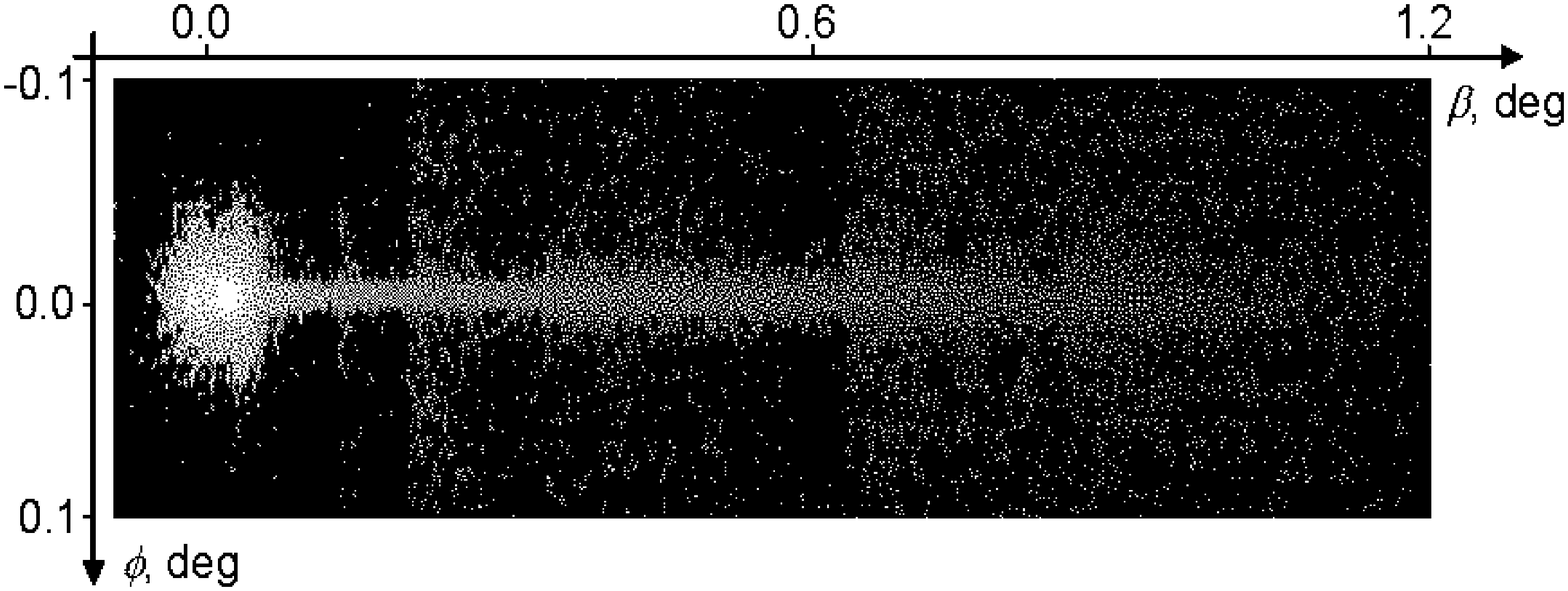, width=0.48\textwidth}

\small {\bf Figure 3.} {\it
Two-dimensional scattering map at $\alpha \approx 0.012^\circ$ on the surface of the silica sol SiO$_2$ with a particle size of $\sim 27$\,nm with the bulk Na concentration $c^+_{Na} \approx 0.06$\,mol/L.
}

\end{figure}

\begin{figure}
\epsfig{file=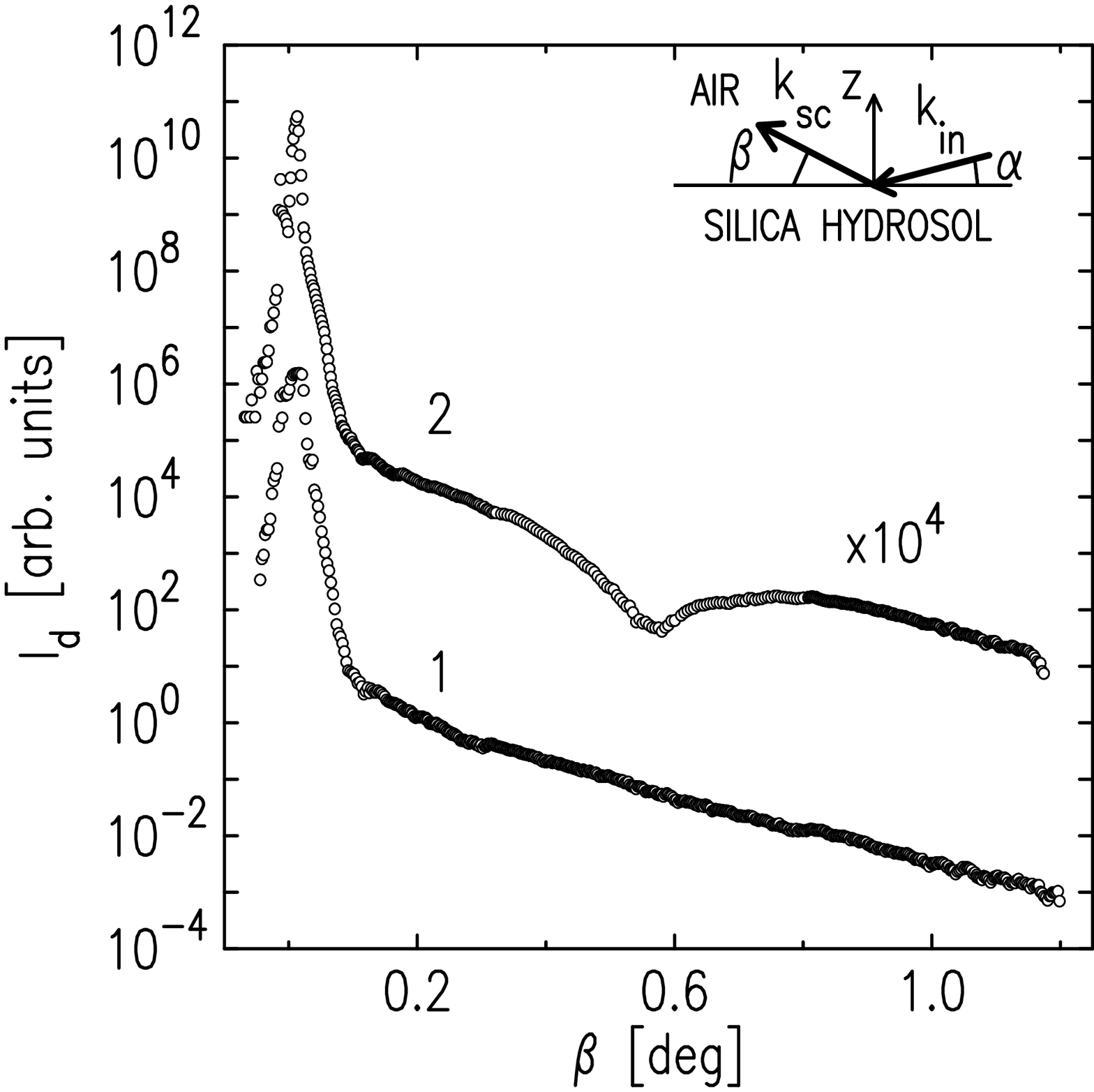, width=0.4\textwidth}

\small {\bf Figure 4.} {\it Diffuse scattering intensity $I_d (\beta)$ for the surfaces of (1) NaOH-stabilized and (2) CsOH-enriched sols at $\alpha \approx 0.012^\circ$. The inset shows the scheme of measurement of diffuse scattering.
}
\end{figure}

In this work, scattering and reflectometry data were jointly analyzed within an iterative approach
described in detail in [7]. Specular reflectivity curves were analyzed within a model-independent method
based on the extrapolation of the asymptotic angular dependence of the specular reflectivity $R(q_z)$ to the region of large $q_z$ values without any prior assumptions on the structure of the sample surface. The polarizability profile $\delta(z)$ thus obtained unambiguously specifies the electron density distribution along the $Oz$ axis $\rho(z)\approx 2\pi\delta(z)/(r_0\lambda^2)$ [22]. Using this approach, we previously studied the structure and formation kinetics of macroscopically flat lipid membranes on the hydrosol surface [23-25].

Diffuse scattering was analyzed within perturbation theory in the function $\zeta(x,y)$ describing the surface relief (roughness) at the medium–air interface. For conformal roughnesses (the function $\zeta(x,y)$ is independent of the distribution of polarizability over the $Oz$ axis and $\langle\zeta(x,y)\rangle = 0$), the two-dimensional distribution of the scattering intensity from the surface (scattering indicatrix) has the form [26,27]
\begin{equation}
\begin{array}{l}
\displaystyle I_d(\mathbf k_{in},\mathbf k_{sc}) = \frac{\displaystyle k_0^4}{\displaystyle (4\pi)^2\sin\alpha}\times
\\ \\
\displaystyle
 \left|
    \int\psi(z, \mathbf k_{in})\psi(z,\mathbf k_{sc})\frac{\displaystyle d\varepsilon}{\displaystyle dz}dz
  \right|^2 \bar C(\nu),
\end{array}
\end{equation}
Here, $\bar C(\nu)$ is the power spectral density function, which is the Fourier transform of the relief autocorrelation function (height-height correlation function) and depends on the length of the spatial frequency vector $\nu= q_y/(2\pi)$, $\varepsilon(z) \approx 1-\delta(z)$ is the distribution of the relative permittivity near the surface, and $\psi(z, \mathbf k)$ is the distribution of the complex amplitude of the wave in the sample, which is numerically found as the solution of the one-dimensional wave equation with the profile $\varepsilon(z)$.

On one hand, since the liquid surface is isotropic, the calculation of the function $\bar C(\nu)$ requires only the indicatrix in the plane of reflection $\phi = 0$). Any prior assumptions on the statistics of its height distribution are not necessary for this calculation. The required information on the distribution $\delta(z)$ can be obtained from the analysis of the angular dependence of the reflectivity curve $R(q_z)$.

\begin{figure}
\epsfig{file=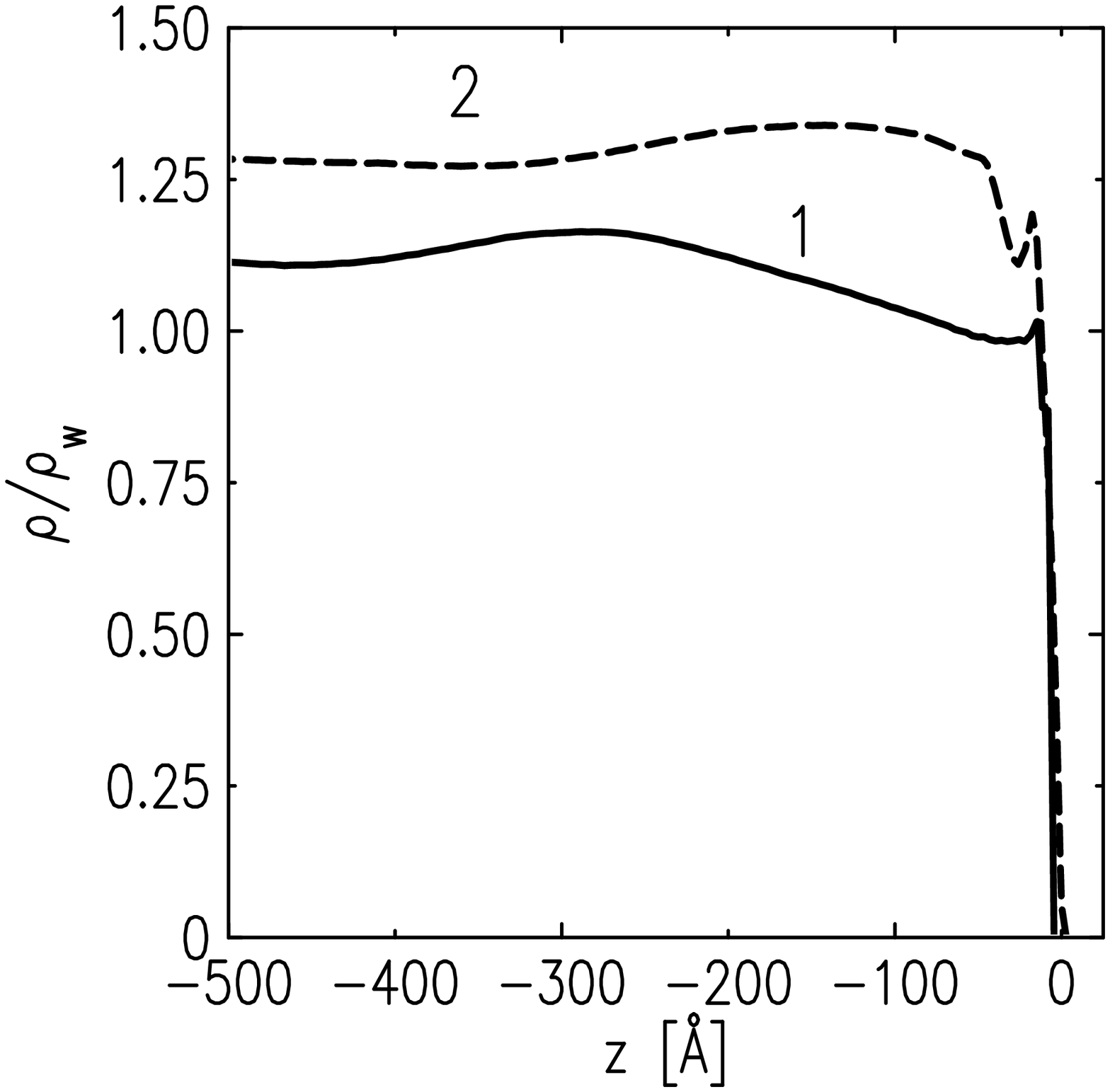, width=0.4\textwidth}

\small {\bf Figure 5.}  {\it Reconstructed distribution profiles $\rho(z)$ normalized to the electron density in water $\rho_w=0.333$\,$e^{-}/\AA^{3}$ for the surfaces of the (solid line 1) NaOH-stabilized and (dashed line 2) CsOH-enriched sols of 27-nm particles. }
\end{figure}

On the other hand, the experimental reflectivity also includes the contribution from diffuse scattering
on roughnesses; for conformal roughnesses, this contribution is given by the Nevot–Croce formula [28]
\begin{equation}
  R(q_z) = R_0(q_z) \cdot \exp\left(-\sigma^2q_z\sqrt{q^2_z - \frac{k_0^2\delta}{4}}\right)
\end{equation}
Here, $\delta \approx 1.1 \cdot 10^{-7}$ for $\lambda=0.1747$\,\AA, $R_0(q_z)$ is the reflectivity from the surface with the profile $\rho(z)$ in the absence of roughness (i.e., $\zeta(x,y)\equiv 0$ at any points $(x,y)$), and
\begin{equation}
  \sigma^2 = \int\limits_{0}^{\infty} \bar C(\nu)d\nu
\end{equation}
is the rms height of roughnesses. Thus, within the used approach, in each iteration step, the surface polarizability profile $\delta(z)$ was first reconstructed with the roughness parameters found at the preceding step and the roughness spectrum $\bar C(\nu)$ was then calculated with this found profile.

Figures 5 and 6 show the reconstructed electron density profiles $\rho(z)$ normalized to the electron density in water $\rho_w=0.333$\,$e^{-}/\AA^{3}$  and roughness spectra $\bar C(\nu)$ for the surfaces of (1) NaOH-stabilized and (2) CsOH-enriched sols of 27-nm particles. It is noteworthy that three iterations appear to be enough to reach a stable solution in which the next iterations for the polarizability profile and roughness spectrum coincide with the preceding iterations for the respective quantities.

Both electron density profiles (see Fig. 5) include a peak with the thickness $d=15 - 20$~\AA{} near the surface; this peak corresponds to the layer of adsorbed ions. The surface density of Na$^+$ ions $\Theta = [\int^0_{-d}\rho(z)dz]/n_e$ ($n_e=10$ is the number of electrons in the Na$^+$ ion) is estimated as $\Theta=(7\pm1)\cdot 10^{18}$\,m$^{-2}$. The electron density in the surface layer of the sol enriched with cesium hydroxide is increased to a value corresponding to the surface density of Cs$^+$ ions ($n_e=54$) $\delta\Theta=(5\pm1)\cdot 10^{18}$\,m$^{-2}$. These values are in agreement with the values
previously obtained within capillary wave models [2,8,13].

\begin{figure}
\epsfig{file=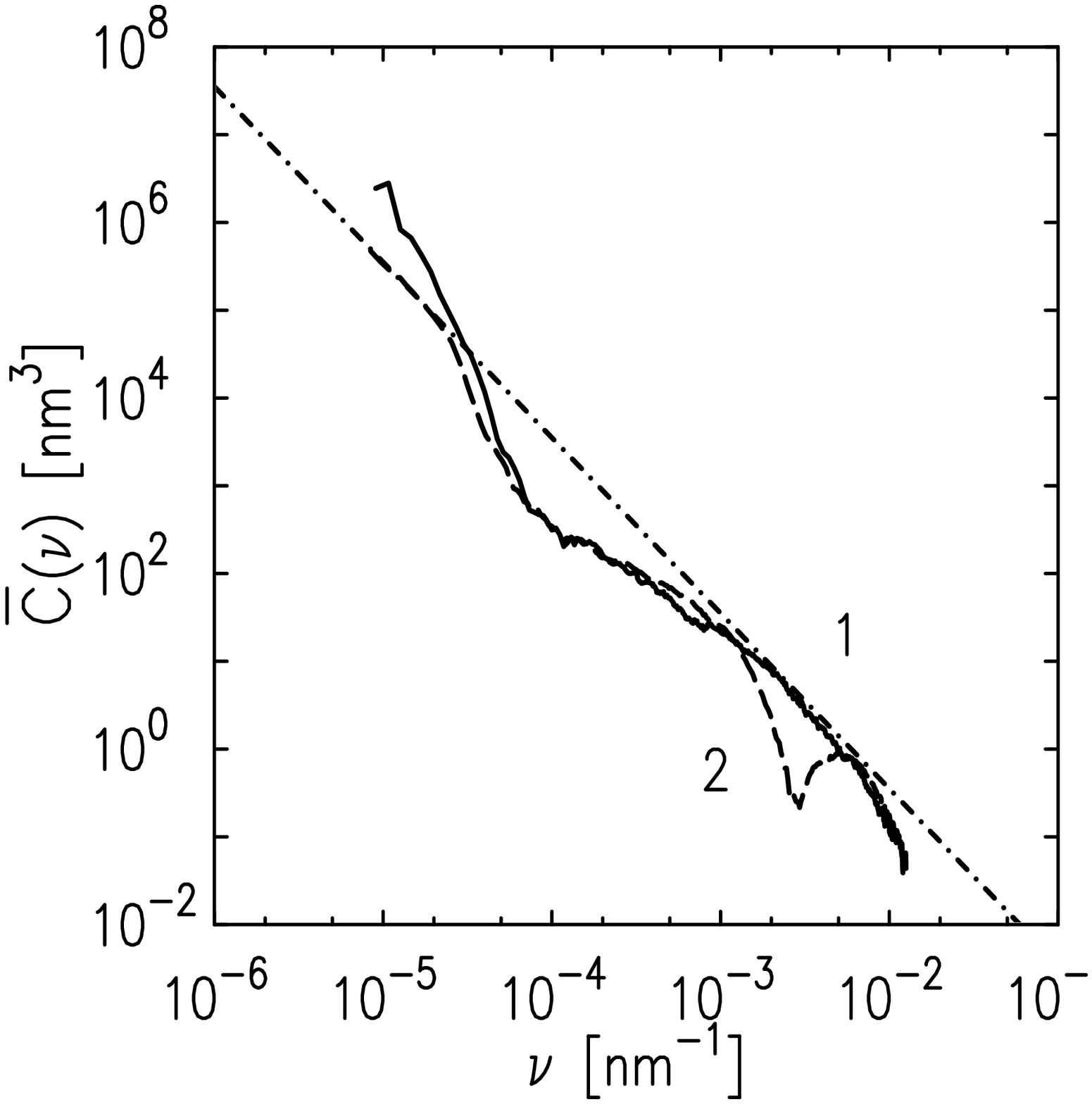, width=0.4\textwidth}

\small {\bf Figure 6.} {\it Roughness spectra $\bar C(\nu)$ according to diffuse scattering data for the surfaces of the (solid line 1) NaOH-stabilized and (dashed line 2) CsOH-enriched sols of 27-nm particles. The dash-dotted line is the theoretical spectrum of the correlation function of capillary wave heights [29]. }
\end{figure}

In turn, the effective roughness $\sigma$ calculated from the functions $\bar C(\nu)$ in the experimentally available range of spatial frequencies $\nu = 10^{-5} - 10^{-2}$\,nm$^{-1}$ (Fig. 6) was $\sigma=(5.7\pm0.3)$\,\AA{} and $\sigma=(2.8\pm0.3)$\,\AA{} for the surfaces of the NaOH-stabilized and Cs-enriched sols, respectively. It is noteworthy that the calculated roughness caused by capillary waves on the silica sol surface is $(2.8\pm0.2)$\AA{} and hardly depends on the ionic composition of the solution [8]. The experimentally available $\nu$ range is limited by the width of the direct beam and maximum $\beta$ value that allows separating surface scattering from scattering in the bulk.

The integral characteristic $\sigma$ of the roughness spectrum of the Cs-enriched sol is in good agreement with the theoretical prediction, whereas this characteristic for the NaOH-stabilized solution is significantly larger than the theoretical prediction. The comparison of the experimental spectrum (dashed line in Fig. 6) with the calculation within the capillary-wave theory [29] (dash-dotted straight line in Fig. 6) shows that the function $\bar C(\nu)$ for the NaOH-stabilized solution at low spatial frequencies $\nu < 10^{-4}$\,nm$^{-1}$ is much larger than capillary wave values. This is possible if the sol surface has two-dimensional spatial inhomogeneities with correlation lengths of $10 - 100$\,$\mu$m, which are not described within the capillary wave formalism, and their distribution deviates from a normal law.

In addition, both dependences $\bar C(\nu)$ exhibit oscillations indicating the interference of diffuse scattering from different layer interfaces of the surface structure [30]. Since the layer of adsorbed ions on the surface has a thickness of $\approx 20$~\AA{} smaller than the characteristic radiation penetration depth in the region of total external reflection $\approx \lambda/(2\pi\alpha_c)\approx 80$\,\AA{} (e.g., see [31]), the observed oscillations in roughness spectra are assumingly due to diffuse scattering at the inner interface between layer 2 and depleted layer 3. To obtain information on the roughness of these hidden interfaces, an additional analysis using an extended procedure described in [7] is necessary, which is beyond the scope of this work.

The mean distance between Na$^+$ cations in layers 1 and 2 is $\sim 5-6$\,\AA{}, which is smaller than the Bjerrum length for univalent ions in the aqueous medium $\sim 7$\,\AA. Many authors believe that ions adsorbed with such a high density, e.g., on the surface of charged particles (macroions) in a colloidal solution, constitute a strongly correlated two-dimensional liquid, where the short-range order is close to a Wigner crystal [32-38].

Alkali metal ions on the silica sol surface can be considered as a heavy, very dense analog of the twodimensional system of "classical" electrons suspended over the surface of some cryogenic insulators (liquid $^3$He, $^4$He, and liquid and solid hydrogen) by electric image forces and external electric field [39]. A solid phase of two-dimensional electrons (Wigner crystal) was previously observed on the surface of liquid helium [40,41]. The temperature in those experiments was much lower than room temperature ($\sim 0.5$\,K), but the experimentally achievable density of the electron gas on the liquid helium surface is a factor of $\sim 10^4$ lower than the density of alkali ions on the silica sol surface. The possibility of appearance of a longrange order on the free surface of the colloidal solution is still incompletely clear because, e.g., the observed translational correlation length between Na$^+$ ions on the surface of the NaOH-stabilized sol is smaller than $\sim 30$\,{\AA}\,[13].

To summarize, we have demonstrated the possibility of using diffuse scattering data measured under the
conditions of total external reflection to obtain information on the statistical properties of the surface of liquid without any prior information on its structure. Electron density profiles and spectra of the height-height correlation function have been reconstructed from experimental data. According to this analysis, the density of suspended Cs$^+$ cations on the surface of the CsOH-enriched sol is $(5\pm1)\cdot 10^{18}$\,m$^{-2}$, which is in good agreement with the data reported in [8]. The revealed deviation from the integral and frequency characteristics of the roughness spectrum of the hydrosol surface from predictions of the capillary-wave theory is of a fundamental character. This deviation is due to the contribution from roughnesses with low spatial frequencies $\nu < 10^{-4}$\,nm$^{-1}$ and to the interference of diffuse scattering from different layer interfaces of the surface structure. The deviation of the integral characteristic $\sigma$ of the spectrum $\bar C(\nu)$ from the values calculated within the capillary-wave theory was previously reported, e.g., for oil–water interfaces in [42,43]. Usually, the spectral features of $\bar C(\nu)$ responsible for this deviation were not indicated in the cited works, where the interpretation was reduced to the assumption of the existence of a certain intrinsic structure.

The experiments on the ID31 beamline at the European Synchrotron Radiation Facility (ESRF, Grenoble, France) were performed within research projects SC-4246 and SC-4461. We are grateful to T. Buslaps (ESRF) for assistance in the use of the ID31 beamline and to Dr. H.\,Reichert and Dr. I.\,V.\,Kozhevnikov for genuine interest and stimulating discussions of the experimental results.

\end{document}